\documentclass{appolb}
\usepackage{graphicx}
\usepackage{amsmath}

\begin{document}
\title{ Study of the Dalitz plot of the $\eta \rightarrow \pi^+ \pi^- \pi^0$ decay \\ with the KLOE detector
%
}
\author{Li Caldeira Balkest\aa{}hl
\address{Uppsala University, Uppsala}
\\
\address{on behalf of the KLOE-2 collaboration}
}
\maketitle
\begin{abstract}
The decay $\eta \rightarrow \pi^+ \pi^- \pi^0$ is studied with the KLOE detector, at the DA$\phi$NE e$^+$e$^-$ collider. Using a data sample corresponding to an integrated luminosity of $1.6$ fb$^{-1}$ a new study of the Dalitz plot is presented. 
\end{abstract}
  
\section{Introduction}
The isospin violating decay $\eta \rightarrow 3\pi$ is a good laboratory to study the light quark mass difference, $m_d - m_u$. According to Sutherland and Bell \cite{sutherland66, sutherland_bell68}, the electromagnetic contributions to this decay are small. Calculations of the electromagnetic corrections in chiral perturbation theory ($\chi$PT) at next-to-leading order (NLO), at the isospin limit, $\mathcal{O}(e^2)$ \cite{baur_kambor_wyler96} and including isospin breaking to leading order, $\mathcal{O}(e^2(m_d - m_u))$ \cite{ditsche_kubis_meissner2009}, show only small effects, confirming Sutherland and Bell's conlcusions. The $\eta \rightarrow 3\pi$ decay thus proceeds mainly by strong interactions and the process can be calculated in $\chi$PT. Up to NLO  the decay rate of  $\eta \rightarrow \pi^+ \pi^- \pi^0$ is related to the quark mass ratio $Q$ as follows:
\[
\Gamma(\eta \rightarrow \pi^+ \pi^- \pi^0) \propto Q^{-4} \qquad Q^2 \equiv \frac{m_s^2 - \hat{m}^2}{m_d^2 - m_u^2} \qquad \hat{m}= \frac{1}{2}(m_d+m_u).
\] 

\begin{figure}[htb]
\centerline{%
\includegraphics[width=.45\textwidth]{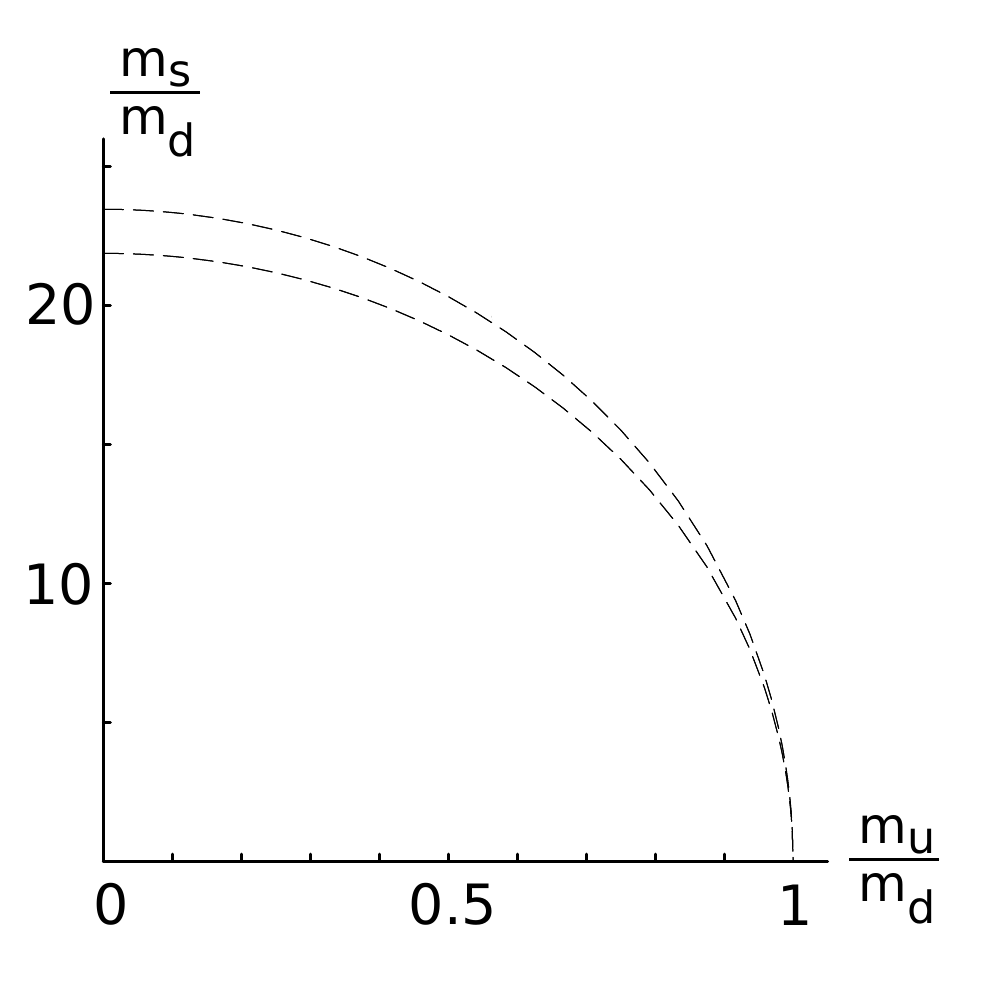}}
\caption{The Leutwyler ellipse shows the contraints for the light quark masses, here for $Q=22.7 \pm 0.8$.}
\label{fig:q2ellips}
\end{figure}

As can be seen from the definition of $Q$, if one neglects the small term $\hat{m}^2/m_s^2$, the formula becomes an ellipse in the $\frac{m_s}{m_d}, \frac{m_u}{m_d}$ plane with major semi-axis $Q$, the Leutwyler ellipse \cite{leutwyler96}, shown in figure \ref{fig:q2ellips}. A precise knowledge of $Q$ is an important constraint for the light quark masses.


At leading order in $\chi$PT, $Q$ can be extracted from meson masses. Assuming Dashen's theorem \cite{dashen69} to fix the electromagnetic contribution to the meson masses gives $Q \equiv Q_D=24.2$. Using $Q_D$ in LO and NLO $\chi$PT calculations results in values of the decay width  $\Gamma(\eta \rightarrow \pi^+ \pi^- \pi^0)$ not in agreement with the experimental value: $\Gamma_{LO}=66$ eV, $\Gamma_{NLO}=160 \pm 50$ eV \cite{gasser_leutwyler85_eta3pi} (updated to  $\Gamma_{NLO}=168 \pm 50$  eV \cite{leutwyler96}) and $\Gamma_{exp}=295 \pm 16$  eV \cite{PDG12}. 

 The big difference between LO and NLO results indicates a slow convergence of the $\chi$PT series and the importance of final state interaction between pions. Instead of using $Q$ to predict the decay width, an accurate theoretical calculation of the decay amplitude together with an accurate experimental value for the decay width can be used to extract $Q$. 

There have been several efforts on improving the decay amplitude calculations. A NNLO $\chi$PT calculation was performed by Bijnens and Ghorbani (in leading order isospin breaking) \cite{bijnens_ghorbani2007}, but at this order many Low Energy Constants must be fixed. Dispersion relations can be used to take into account the $\pi \pi$ interactions, and two recent different methods have been used to improve the NLO $\chi$PT result \cite{kampf_knecht_novotny_zdrahal2011,colangelo_lanz_leutwyler_passemar2011}.
These promising methods can use as input the experimental Dalitz plot density information for the charged $\eta \to \pi^+ \pi^- \pi^0$ channel. Their predictions provide a precise determination of $Q$, and also provide an important consistency check, namely to get a reasonable agreement with the experimental results for the neutral channel $\eta \to 3\pi^0$. As one of the main sources of uncertainty on $Q$ comes from the uncertainty on the input parameters, it is crucial to precisely measure the density distribution of the $\eta \to \pi^+ \pi^- \pi^0$ Dalitz plot.

\subsection{Dalitz plot for the charged decay}
Only few experiments since the 70s have measured the  $\eta \rightarrow \pi^+ \pi^- \pi^0$ Dalitz plot distribution. The results for the extracted Dalitz plot parameters are shown in Table \ref{tab:dpexp}. This Dalitz plot is usually presentend using the $X$ and $Y$ variables:
\[
X= \sqrt{3} \frac{T_{\pi^+} - T_{\pi^-}}{Q_\eta} = \frac{\sqrt{3}}{2 m_\eta Q_\eta}(u-t)
\]
\[ Y= \frac{3T_{\pi^0}}{Q_\eta} -1 =  \frac{3}{2 m_\eta Q_\eta} \left[ \left(  m_\eta - m_{\pi^0} \right)^2 - s \right] -1 
\]
\[ Q_\eta =T_{\pi^+} +  T_{\pi^-} + T_{\pi^0} = m_\eta - 2m_{\pi^+} -m_{\pi^0}
\]
\label{sec:dpintro}
where $T$ is the kinetic energy of the different pions in the $\eta$ rest frame and $s,t,u$ are the usual Mandelstam variables.
The Dalitz plot parameters $a,b,c, \ldots $, presented in the table, are defined by a polynomial expansion of the amplitude squared in $X$ and $Y$, $|A(X,Y)|^2 \simeq N(1+aY+bY^2+cX+dX^2+eXY + fY^3+ gX^2Y + hX Y^2 + lX^3)$ \footnote{The parameters $c,e,h$ and $l$ would signal a violation of charge conjugation invariance and are therefore usually fixed to zero in the fit of the Dalitz plot distribution.}. 

As can be seen in Table  \ref{tab:dpexp}, there is some tension between the two most recent, high statistics experiments (KLOE \cite{kloe2008} and WASA \cite{patrik2014}), and more data is needed to provide a clearer picture. A new KLOE analysis, using an independent data set, a better understanding of the detector and improved analysis methods aims to provide a new measurement with reduced systematic uncertainties. The status and methodology of this analysis is presented here. 

\begin{table}
\begin{tabular}{l@{}| l l l l}
Experiment & $-a$ & $b$ & $d$ & $f$ \\
\hline
Gormley(70) \cite{gormley70} & 1.17(2) & 0.21(3) & 0.06(4) & - \\
Layter(73) \cite{layter73} & 1.080(14) & 0.03(3) & 0.05(3) & - \\
CBarrel(98) \cite{CBarrel98} & 1.22(7) & 0.22(11) & 0.06(fixed) & -\\
KLOE(08) \cite{kloe2008} & $1.090(5)(^{+19}_{-8})$ & $0.124(6)(10)$ & $0.057(6)(^{+7}_{-16})$ & 0.14(1)(2) \\
WASA(14) \cite{patrik2014} & 1.144(18) & 0.219(19)(47)  & 0.086(18)(15)& 0.115(37)
\end{tabular}
\caption{Experimental results for Dalitz plot parameters of $\eta \rightarrow \pi^+ \pi^- \pi^0$.\label{tab:dpexp}}
\end{table}

\section{Signal event selection}
The analysis is performed using  $1.6$ fb$^{-1}$ of data collected with the KLOE detector in 2004-2005 (the previous analysis was based on $450$ pb$^{-1}$) . The KLOE detector is located in the DA$\phi$NE $\phi$-factory, and the $\eta$ meson is produced via the radiative decay $\phi \rightarrow \eta \gamma_\phi$. For the $\eta \rightarrow \pi^+ \pi^- \pi^0$ Dalitz plot analysis, with $\pi^0 \rightarrow \gamma \gamma$, we thus have two charged tracks and three photons in the final state. The event selection requires three clusters in the calorimeter \cite{KLOEEMCNIM} consistent with photons\footnote{Not connected to a track and within the expected time window for a massless particle.} and two tracks with opposite curvature in the drift chamber \cite{KLOEDCNIM}. 

\begin{figure}[htb]
\centerline{%
\includegraphics[width=.93\textwidth]{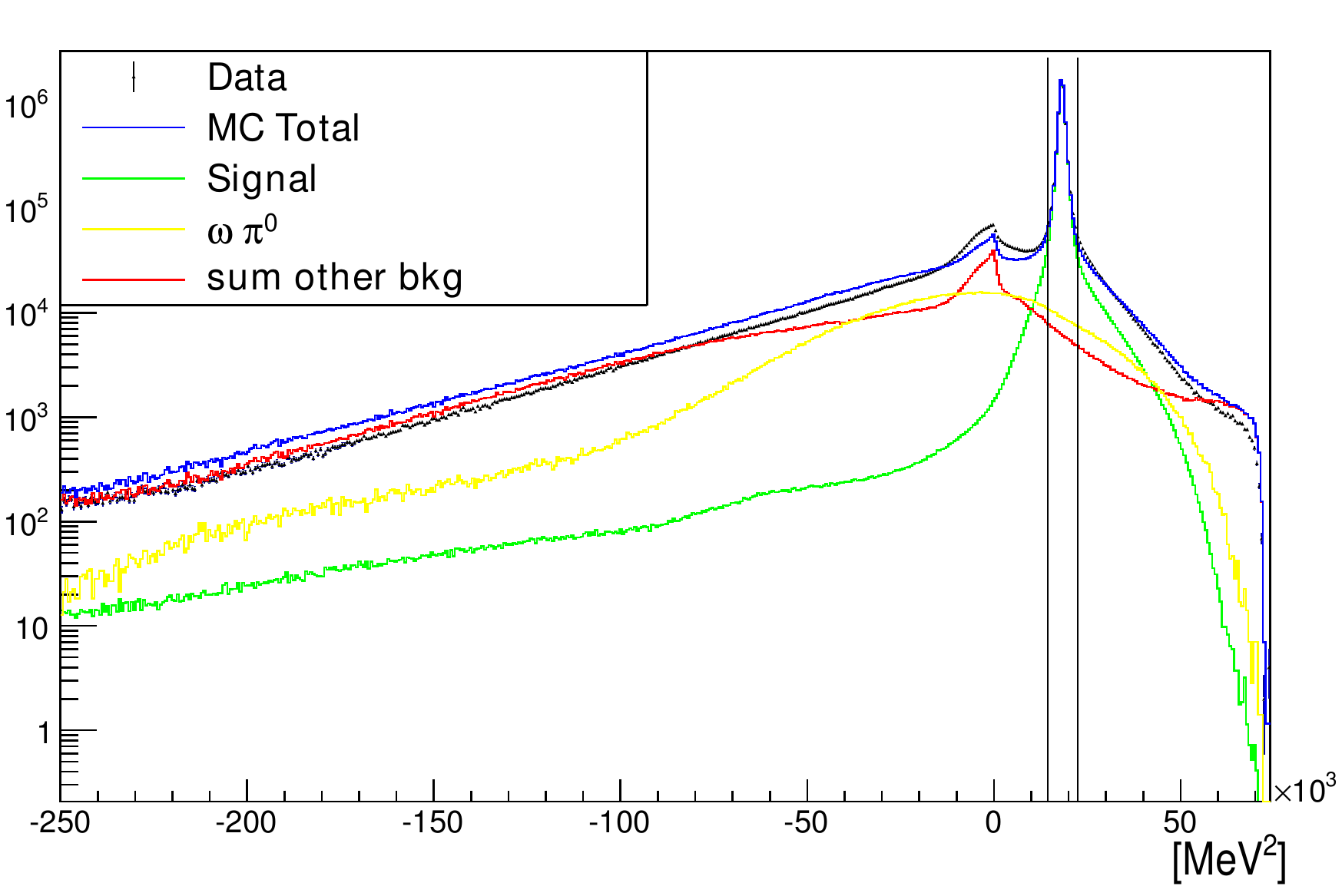}}

\centerline{%
\includegraphics[width=.93\textwidth]{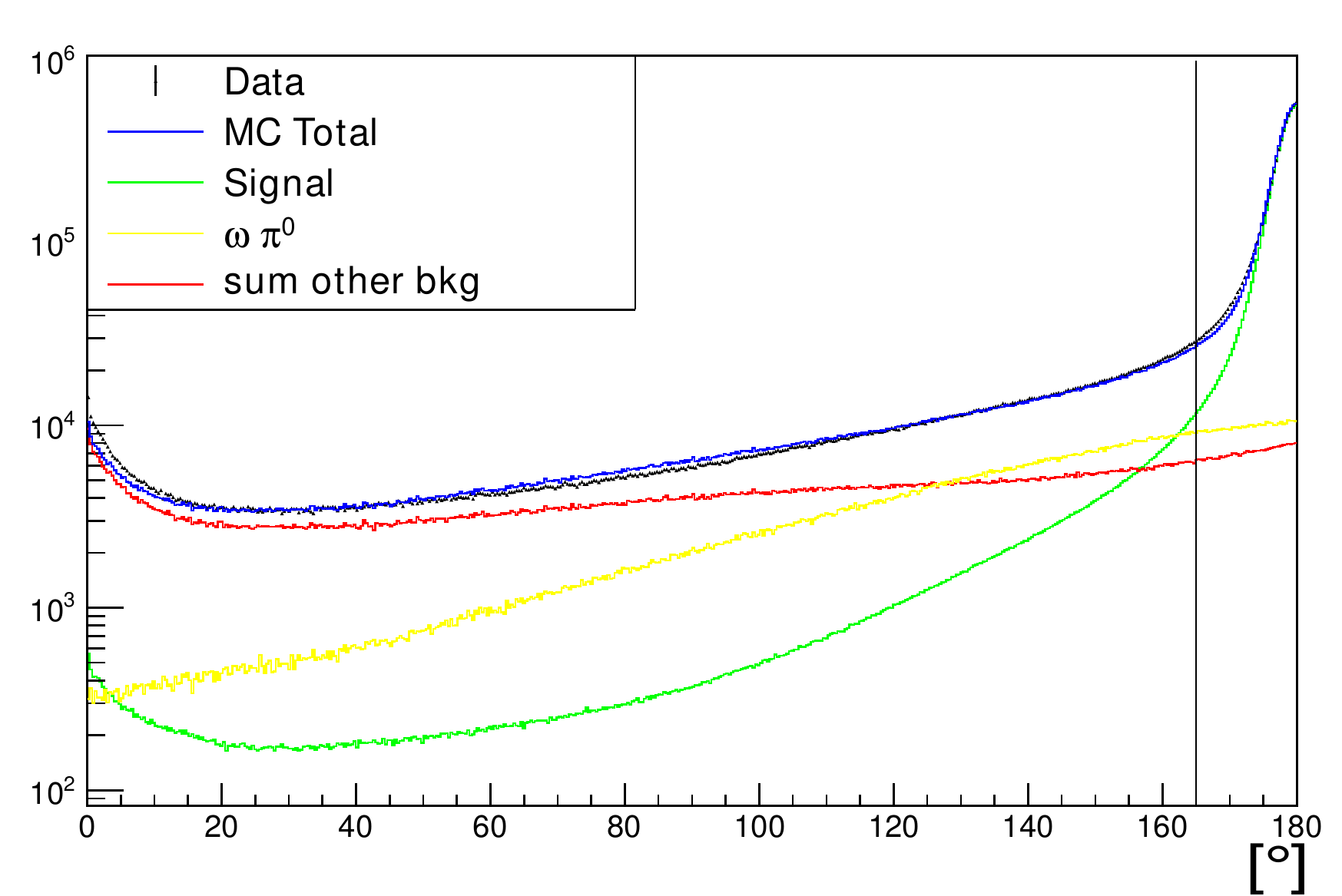}}
\caption{Data-Monte Carlo comparison for the missing mass squared (top) and opening angle between $\pi^0$ photons in the $\pi^0$ rest frame (bottom). The vertical lines show the selection cuts.}
\label{fig:datamc}
\end{figure}

To improve the signal to background ratio several cuts are applied: on the angle between the tracks and the photons; on the time-of-flight used for particle identification; on the angle between the $\pi^0$ decay photons in the $\pi^0$ rest frame ($\Theta_{\pi^0}$); and on the missing mass $MM(\phi - \pi^+ - \pi^- - \gamma_\phi)$. The two last cuts are depicted in Figure \ref{fig:datamc}, where both data and Monte Carlo simulation are shown. These histograms are used to fix the scaling factors used for the Monte Carlo background, and, as can be seen in Figure \ref{fig:datamc}, a  good data-Monte Carlo agreement is obtained. After reconstruction and cleanig cuts the signal efficiency is $37.6\%$ with a $1\%$ background contamination, evaluated from Monte Carlo. 

\section{Fit of the Dalitz plot}
\begin{figure}[htb]
\centerline{%
\includegraphics[width=.85\textwidth]{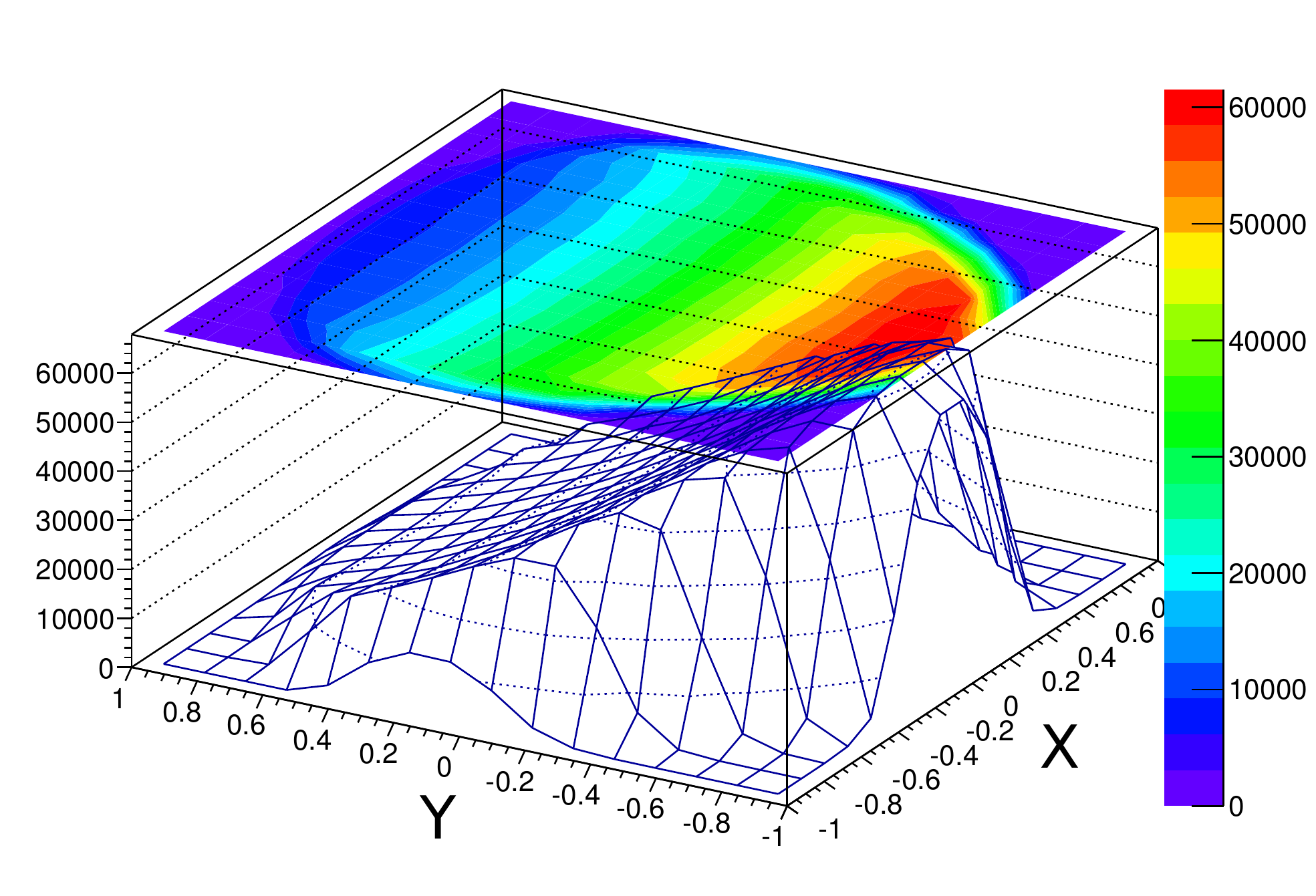}}
\caption{Experimental Dalitz plot of $\eta \rightarrow \pi^+ \pi^- \pi^0$.}
\label{fig:dp}
\end{figure}

The Dalitz plot parameters are obtained by a fit of  the parametrization shown in section \ref{sec:dpintro} to the measured Dalitz plot density, by minimizing:
\[\chi^2 = \sum_{i=1}^{Nbins} \left( \frac{N_i - \sum_{j=1}^{Nbins} \epsilon_j S_{ij} N_{theory,j}}{\sigma_i}    \right)^2,
\]
where  $ N_i$ is the background subtracted data content in Dalitz plot bin i, $\epsilon_j$ is the acceptance of bin j, $S_{ij}$ is the smearing matrix from bin j to bin i,  $N_{theory,j} = \int |A(X,Y)|^2 dXdY$ (inside the phase space) is calculated with Monte Carlo integration for each bin, and $\sigma_i$, the error in bin i, includes the error in $N_i$ and in $\epsilon_j S_{ij}$. Figure \ref{fig:dp} shows the experimental Dalitz plot used.
 
From a detailed study both on data and Monte Carlo, fixing to zero the $c$ and $e$ parameters (and keeping the amplitude squared expansion up to the $f$ term), it is found that the statistical uncertainty on the $a,b,d$ and $f$ parameters can be reduced by about a factor of two with respect to the previous KLOE result (see Table \ref{tab:dpexp}), while improving also the systematic uncertainties. 

The inclusion in the fit of the $g$ parameter would also be possible and its impact on the stability of the result is currently under study.

In the near future, a further significant improvement of both statistical and systematic uncertainties of the Dalitz plot parameters will be possible at KLOE-2 \cite{kloe2physics}, thanks to the luminosity upgrade of DA$\phi$NE and to the better quality of reconstructed data using new detectors: inner tracker \cite{kloeIT}, crystal calorimeters (CCALT) \cite{kloeCCALT} and a tile calorimeter (QCALT) \cite{kloeQCALT} of the upgraded KLOE detector. 

\section*{Acknowledgements}
We warmly thank our former KLOE colleagues for the access to the data collected during the KLOE data taking campaign.
We thank the DA$\Phi$NE team for their efforts in maintaining low background running conditions and their collaboration during all data taking. We want to thank our technical staff: 
G.F. Fortugno and F. Sborzacchi for their dedication in ensuring efficient operation of the KLOE computing facilities; 
M. Anelli for his continuous attention to the gas system and detector safety; 
A. Balla, M. Gatta, G. Corradi and G. Papalino for electronics maintenance; 
M. Santoni, G. Paoluzzi and R. Rosellini for general detector support; 
C. Piscitelli for his help during major maintenance periods. 
This work was supported in part by the EU Integrated Infrastructure Initiative Hadron Physics Project under contract number RII3-CT- 2004-506078; by the European Commission under the 7th Framework Programme through the `Research Infrastructures' action of the `Capacities' Programme, Call: FP7-INFRASTRUCTURES-2008-1, Grant Agreement No. 227431; by the Polish National Science Centre through the Grants No. 
DEC-2011/03/N/ST2/02641, 
2011/01/D/ST2/00748,
2011/03/N/ST2/02652,
2013/08/M/ST2/00323,
and by the Foundation for Polish Science through the MPD programme and the project HOMING PLUS BIS/2011-4/3.

\bibliographystyle{lisstyle}
\bibliography{eta3pi}

\end{document}